\documentstyle[prd,aps,twocolumn,floats]{revtex}
\renewcommand{\epsilon}{\varepsilon}
\begin{document}
\twocolumn[\hsize\textwidth\columnwidth\hsize\csname
@twocolumnfalse\endcsname
\title{Cosmological thermodynamics and  deflationary gas universe}
\author{Winfried Zimdahl\footnote{Electronic address: winfried.zimdahl@uni-konstanz.de}}
\address{Fakult\"at f\"ur Physik, Universit\"at Konstanz, PF 5560 M678
D-78457 Konstanz, Germany}
\author{Alexander B. Balakin
\footnote{Electronic address: dulkyn@bancorp.ru}} 
\address{Fakult\"at f\"ur Physik, Universit\"at Konstanz, PF 5560 M678
D-78457 Konstanz, Germany\\
and 
Department of General Relativity and Gravitation,  
Kazan State University, 420008 Kazan, Russia\thanks{Present address}}
\date{\today}
\maketitle
\begin{abstract}
We establish a general thermodynamic scheme for cosmic fluids with internal self-interactions and discuss equilibrium and non-equilibrium aspects 
of such systems in connection with (generalized) symmetry properties of the cosmological dynamics. 
As an example we construct an exactly solvable gas dynamical model of a ``deflationary'' transition from an initial de Sitter phase to a subsequent Friedmann-Lema\^{\i}tre-Robertson-Walker 
(FLRW) period. We demonstrate that this dynamics represents a manifestation of a conformal symmetry of an ``optical'' metric, characterized by a specific effective refraction index of the cosmic medium. 
\end{abstract}  
\vspace{1.5cm}
]

\section{Introduction}

A comprehensive picture of the physics of the early universe crucially relies on an adequate understanding of thermodynamic aspects of the cosmological evolution \cite{KoTu,Boe}. 
Of particular interest are periods during which the cosmic substratum was out of equilibrium. 
It is generally believed that essential properties of the presently observed universe may be traced back to specific non-equilibrium processes in early cosmology. 
A detailed and transparent description of out-of equilibrium phases is therefore a basic challenge for any cosmological model. 
On the other hand, the natural starting point for thermodynamical studies is a characterization of possible equilibrium states of the system under consideration. 
It is well known that equilibrium states in the expanding universe are only possible under certain  conditions. 
For a simple gas model e.g., the cosmic matter has to obey exactly an equation of state $p=\rho /3$ 
($p$ is the pressure and $\rho $ is the energy density) \cite{Stew,Ehl}. 
This equilibrium condition is equivalent to a symmetry requirement for the cosmological dynamics: 
The quantity $u ^{a}/T$, where $u ^{a}$ is the four velocity of the medium and $T$ is its temperature, has to be a conformal Killing vector (CKV). 
For deviations from $p=\rho /3$ the conformal symmetry is violated and no equilibrium is possible. 
As long as the deviations from equilibrium are small, there exist well-established approaches and techniques to investigate the corresponding dynamics (see, e.g. \cite{Stew,Ehl,IS,Groot}). 
For larger deviations the situation is less clear and new concepts are required. 

In this paper we focus on a specific class of non-equilibrium states which are characterized by a simple modification of the standard equilibrium and symmetry conditions for relativistic media. 
We demonstrate how the standard equilibrium of relativistic gas dynamics may be generalized such that ultrarelativistic matter is no longer a singular case but a well-defined limit of a more general equilibrium configuration. 
This generalization is achieved by giving up the simple dilute gas approximation and admitting additional internal interactions inside the many-particle system. 
As long as these interactions are particle number preserving the corresponding states are real equilibrium states. 
The step away from equilibrium is then performed by including internal interactions which do {\it not} preserve the particle number. 
This procedure, although formally rather straightforward within the presented formalism, is physically essentially non-trivial since it introduces states with 
(not necessarily small) entropy production. 
It is a remarkable feature of this kind of non-equilibrium that, irrespective of the amount of entropy production, it is characterized by an equilibrium distribution (``generalized'' equilibrium) of the microscopic particles and 
is accessible by an analytic treatment. 
Furthermore, there exists a threshold for the entropy production in homogeneous and isotropic models beyond which such type of states are only possible in a universe with accelerated expansion. 
This represents a surprizing link between (generalized) equilibrium properties of cosmic matter and the expansion behaviour of the universe. 

The present paper generalizes and extends previous investigations
\cite{ZTP,ZiBa1,ZiBa2,ZBGRG} on specific aspects of generalized equilibrium configurations in a cosmological context.  
Our first purpose is to provide a general characterization of all those (equilibrium and non-equilibrium) states  of a gaseous cosmological fluid  which on the microscopic level are governed by an equilibrium distribution function. 
The second purpose is to establish on this basis an exactly solvable 
homogeneous and isotropic gas dynamical model according to which the universe starts in a phase of exponential expansion and subsequently smoothly evolves towards a standard radiation dominated FLRW phase. 
This ``deflationary'' \cite{Barrow} transition is shown to be the manifestation of a specific non-equilibrium configuration of the cosmic medium. 
We determine those forces on the microscopic constituents of the cosmic matter which underly this dynamics. 
Moreover, we demonstrate that a non-equilibrium transition of this kind represents a conformal symmetry of an ``optical'' metric 
\cite{Gordon,Ehlopt,Perlick} 
with an initially large value of the ``refraction'' index of the universe which continously decreases to unity as the FLRW phase is approached. 

The paper is organized as follows. 
In Sec. II we establish the general thermo-hydrodynamical framework for a relativistic fluid with particle production and effective viscous pressure. 
This scheme essentially relies on a consistency requirement for the temperature evolution of the medium which relates symmetry aspects with those of thermodynamics. 
A corresponding microscopic realization is given in section III,  
where we study the kinetic theory of a gas of self-interacting particles. 
This encompasses a determination of the effective self-interacting force by generalized equilibrium requirements and a derivation of corresponding transport equations, followed by an investigation of the microscopic particle motion in such a system. 
In Sec. IV we demontrate how a specific self-interaction in a homogeneous and isotropic cosmic medium implies a deflationary \cite{Barrow} transition from an initial de Sitter phase to a subsequent FLRW period. 
We determine an optical metric with respect to wich the quantity $u ^{a}/T$ is a CKV for the entire transition process. 
An equivalent two-component description is given in Sec. V, where the deflationary dynamics is interpreted in terms of a decaying vacuum. 
Section VI sums up the main features and conclusions of the paper. 
Units have been chosen so that $c = k_{B} =  \hbar =1$. 

\section{Thermo-hydrodynamic approach}

We assume the cosmic substratum to be modeled by a fluid with the energy-momentum tensor 
\begin{equation}
T _{\left(eff \right)} ^{ik} = \rho  u ^{i}u ^{k} + P h ^{ik}\ ,
\label{1}
\end{equation}
where $\rho $ is the energy density measured by an observer comoving with the fluid four-velocity $u ^{i}$, normalized by $u ^{a}u _{a} = -1$, and   
$h _{ik} \equiv  g _{ik} + u _{i}u _{k}$. 
The total pressure $P$ is the sum  
\begin{equation}
P = p + \Pi 
\label{2}
\end{equation}
of an equilibrium part $p>0$ and an effective non-equilibrium part $\Pi<0$  which is connected with entropy production. 
Within the Eckart frame the particle number flow vector $N ^{i}$ is 
given by 
\begin{equation}
N ^{i} = n u ^{i}
\label{3}
\end{equation}
where $n$ is the particle number density. 
While we require local energy-momentum conservation 
$T ^{ik}_{\left(eff \right);k} = 0$ 
in accordance with the integrability conditions of Einstein's equations, 
we will not assume the fluid particle number to be conserved. Instead, 
we admit a source term $\Gamma $ in the corresponding balance equation 
which describes the rate of change of the fluid particles:
\begin{equation}
N ^{i}_{;i} = \dot{n} + \Theta n = n \Gamma \ ,
\label{4}
\end{equation}
where $\Theta \equiv  u ^{i}_{;i}$ is the fluid expansion and 
$\dot{n} \equiv  n _{,i}u ^{i}$. 
For $\Gamma > 0$ we have particle creation, for $\Gamma < 0$ particles are annihilated. 
The particle number may change either due to internal reactions within the medium \cite{WZ1,ZPM} or due to particle production in strongly varying gravitational fields 
\cite{Zel,Mur,Hu,BiDa,Turok,Barr,Ford}. 
The production rate $\Gamma $  is an input quantity in a phenomenological description which in principle has 
to be calculated from the underlying microphysics. 
In the general part of this paper $\Gamma $ will be a free parameter. 
In the subsequent application a specific dependence on the Hubble rate of an homogeneous and isotropc universe is considered. 

Energy-momentum conservation is equivalent to 
\begin{equation}
\dot{\rho} +\Theta\left(\rho + p + \Pi\right) = 0\ ,
\label{5}
\end{equation}
and 
\begin{equation}
\left(\rho + p + \Pi  \right)\dot{u}_{a} + \nabla  _{a}\left(p + \Pi  \right) 
= 0 \ ,
\label{6}
\end{equation}
where $\dot{u}_{a} \equiv  u _{a;c}u ^{c}$ is the four-acceleration, 
and 
$\nabla  _{a}p \equiv  h _{a}^{b}p _{,b}$ is the covariantly defined spatial derivative.  
With the help of the Gibbs equation (see, e.g., \cite{Groot})
\begin{equation}
T \mbox{d}s = \mbox{d} \frac{\rho }{n} + p \mbox{d}\frac{1}{n}\ ,
\label{7}
\end{equation}
where $s$ is the entropy per particle and $T$ is the equilibrium temperature, we obtain 
\begin{equation}
n T \dot{s} = \dot{\rho } - \left(\rho + p \right)
\frac{\dot{n}}{n}\ .
\label{8}
\end{equation}
Using here the balances (\ref{4})  and (\ref{5})  yields
\begin{equation}
n T \dot{s} = - \Theta \Pi - \left(\rho + p \right)\Gamma \ .
\label{9}
\end{equation}
From the Gibbs-Duhem relation (see, e.g., \cite{Groot})
\begin{equation}
\mbox{d} p = \left(\rho + p \right)\frac{\mbox{d} T}{T} 
+ n T \mbox{d} \left(\frac{\mu }{T} \right)\ ,
\label{10}
\end{equation}
where $\mu $ is the chemical potential,  
it follows that 
\begin{equation}
\left(\frac{\mu }{T}\right)^{^{\displaystyle \cdot}} 
= \frac{\dot{p}}{nT} 
- \frac{\rho + p}{nT}\frac{\dot{T}}{T}\ .
\label{11}
\end{equation}
We assume  equations of state in the general form 
\begin{equation}
p = p \left(n,T \right)\ ,\ \ \ \ \ \ \ \ \ 
\rho = \rho \left(n,T \right)\ , 
\label{12}
\end{equation}
i.e., particle number density and temperature are taken as the independent thermodynamical variables. 
Differentiating the latter relation and using 
the balances (\ref{4}) and  
(\ref{5}) provides us with an evolution law for the temperature: 
\begin{equation}
\frac{\dot{T}}{T} = - \left(\Theta - \Gamma  \right) 
\frac{\partial p}{\partial \rho } 
+ \frac{n \dot{s}}{\partial \rho / \partial T}\ ,
\label{13}
\end{equation}
where the abbreviations
\[
\frac{\partial{p}}{\partial{\rho }} \equiv  
\frac{\left(\partial p/ \partial T \right)_{n}}
{\left(\partial \rho / \partial T \right)_{n}} \ ,
\ \ \ \ \ \ \ \ 
\frac{\partial{\rho }}{\partial{T}} \equiv  
\left(\frac{\partial \rho }{\partial T} \right)_{n}\ ,
\]
have been used, as well as the general relation 
\[
\frac{\partial{\rho }}{\partial{n}} = \frac{\rho + p}{n} 
- \frac{T}{n}\frac{\partial{p}}{\partial{T}}\ ,
\] 
which follows from the fact that the entropy is a state function, i.e., 
$\partial ^{2}s/ \partial n \partial T 
= \partial ^{2}s/ \partial T \partial n$. 
With the help of Eqs. (\ref{12}), (\ref{4}), (\ref{8})  and (\ref{13}) 
$\dot{p}$  may generally be written as     
\begin{equation}
\dot{p} = c _{s}^{2} \dot{\rho } 
+ nT \dot{s} \left[\frac{\partial{p}}{\partial{\rho }} 
- c _{s}^{2}   \right]\ ,
\label{14}
\end{equation}
where 
\begin{equation}
c _{s}^{2}  = \left(\frac{\partial{p}}{\partial{\rho }} \right)_{ad} 
= \frac{n}{\rho + p}\frac{\partial{p}}{\partial{n}} 
+ \frac{T}{\rho + p} 
\frac{\left(\partial p / \partial T \right)^{2}}
{\partial \rho / \partial T}
\label{15}
\end{equation}
is the square of the adiabatic sound velocity $c _{s}$ [see, e.g. \cite{Weinberg}].  
Using Eq. (\ref{14})  in Eq. (\ref{11})  yields 
\begin{eqnarray}
\left(\frac{\mu }{T} \right)^{^{\displaystyle \cdot}} 
&=& \frac{\rho + p + \Pi }{nT}
\left(\frac{\partial{p}}{\partial{\rho }} - c _{s}^{2} \right)\Theta 
\nonumber\\
&&+ \left[2 \frac{\partial{p}}{\partial{\rho }} - c _{s}^{2}   
- \frac{\rho + p}{T \partial \rho / \partial T}\right]\dot{s}\ .
\label{16}
\end{eqnarray}
To obtain an expression for the spatial variation of $\mu /T$ we combine the Gibbs-Duhem equation (\ref{10})  with the momentum balance (\ref{6}).  
The result is 
\begin{equation}
\nabla  _{a}\left(\frac{\mu }{T} \right) 
= - \frac{\rho + p}{nT}\left(\dot{u}_{a} + \frac{\nabla _{a}T}{T} \right) 
- \frac{1}{nT}\left(\Pi \dot{u} _{a} + \nabla  _{a}\Pi  \right)\ .
\label{17}
\end{equation}
Together with the particle number balance (\ref{4})  the entropy flow vector 
$S ^{a} = ns u ^{a}$ gives rise to the following expression for the entropy production density: 
\begin{equation}
S ^{a} = ns u ^{a} 
\quad\Rightarrow\quad 
S ^{a}_{;a}=ns \Gamma  + n \dot{s}\ .
\label{18}
\end{equation}
It is useful to rewrite the latter as  
\begin{eqnarray}
S ^{i}_{;i} - n s \Gamma  &=&  
- \frac{\rho + p}{T}\Gamma 
- \frac{1}{2}\Pi h ^{ik}
\left[\left(\frac{u _{i}}{T}\right)_{;k} 
+ \left(\frac{u _{k}}{T}\right)_{;i}\right]        \nonumber\\
&=& - \frac{\rho + p}{T}\Gamma 
- \frac{1}{2}\Pi h ^{ik}
\pounds _{\xi } g _{ik}\ ,
\label{19}
\end{eqnarray}
where $\pounds _{\xi }$ denotes the Lie derivative with respect to the temperature vector 
$\xi ^{a} \equiv  u ^{a}/T$. 
The equivalence between $\dot{s}$ in Eq. (\ref{9})  and the right-hand side of the last relation may be checked by direct calculation. 
The purpose of this rewriting is to relate  
the entropy production density to the Lie derivative of the metric 
which, in turn,  may be used to characterize symmetries of the spacetime. 

As any symmetric tensor, the Lie derivative may be split 
into contributions parallel and perpendicular to the four-velocity:
\begin{equation}
\pounds _{\xi } g _{ik} = {\rm 2} A u _{i}u _{k} + B _{i}u _{k} + B _{k}u _{i} 
+ {\rm 2} \phi h _{ik} + b _{ik}  \ ,
\label{20}
\end{equation}
where 
\begin{equation}
A = \frac{1}{2}u ^{i}u ^{k}\pounds _{\xi } g _{ik} 
= \frac{\dot{T}}{T ^{{\rm 2}}}\ ,
\label{21}
\end{equation}
\begin{equation}
B _{m} = - h ^{i}_{m}u ^{k}\pounds _{\xi } g _{ik} 
= - \frac{{\rm 1}}{T}\left[\dot{u}_{m} + \frac{\nabla  _{m}T}{T} \right]\ ,
\label{22}
\end{equation}
\begin{equation}
\phi = \frac{1}{6}h ^{ik}\pounds _{\xi } g _{ik} 
=  \frac{{\rm 1}}{{\rm 3}}
\frac{{\rm \Theta } }{T}\ ,
\label{23}
\end{equation}
and 
\begin{eqnarray}
b _{ab} &=& \left[h _{a}^{i}h _{b}^{k} - \frac{1}{3}h _{ab}h ^{ik} \right]
\pounds _{\xi } g _{ik}\nonumber\\
&=& \frac{{\rm 1}}{T}\left[\nabla  _{b}u _{a} + \nabla  _{a}u _{b} 
- \frac{{\rm 2}}{{\rm 3}}{\rm \Theta } h _{ab}\right] 
\equiv  \frac{{\rm 2} \sigma _{ab}}{T}\ , 
\label{24}
\end{eqnarray}
the quantity $\sigma _{ab}$ being the shear. 

It is crucial for the following considerations that the temperature behavior of the fluid is governed by the general thermodynamics of the fluid according to the law (\ref{13})  and, on the other hand, by the projection 
(\ref{21}) of the Lie derivative. 
Consistency between the expressions (\ref{13})  and 
(\ref{21})  requires 
\begin{equation}
A = - \left(1 - \frac{\Gamma }{\Theta } \right)3 \phi \frac{\partial{p}}{\partial{\rho }} 
+ \frac{1}{2}\frac{n \dot{s}}{T \partial \rho / \partial T}\ .
\label{25}
\end{equation}
All relations considered so far, including the last one, become especially simple 
for $\dot{s}=0$. 
Under this condition the entropy production is entirely determined by the particle production rate, i.e.,  $S ^{a}_{;a}=ns\Gamma $. 
For a conserved particle number the entropy production vanishes. 
Only due to the enlargement of its phase space the system increases its  entropy content, i.e., each newly created particle contributes with its 
(equilibrium) entropy to the total entropy of the system. 
There are, however, no dissipative processes  connected with a change in the entropy per particle. 
Except for a change in the particle number (of whatsoever reason) the system behaves as a perfect fluid. 
According to Eq. (\ref{9})  the condition $\dot{s}=0$ establishes a simple relation between 
the non-equilibrium pressure $\Pi $ and the particle creation rate 
$\Gamma $ \cite{Prig,Calv}:
\begin{equation}
\dot{s} = 0 \quad\Rightarrow\quad
\Pi = - \left(\rho + p \right)\frac{\Gamma }{\Theta }\ .
\label{26}
\end{equation}
With such type of viscous pressure the energy balance (\ref{5}) is equivalent to 
\begin{equation}
\dot{\rho } + \Theta \left(\rho + p \right) = \left(\rho + p \right)\Gamma \ .
\label{27}
\end{equation}
Consequently, the system  may be understood as a perfect fluid with source terms. 
The microscopic counterpart (see below) of this statement is the fact that corresponding states for a gas are described by an equilibrium distribution function. 

With relation (\ref{25})  and for $\dot{s}=0$,   the general structure of the Lie derivative becomes 
\begin{eqnarray}
\pounds _{\xi } g _{ik}  &=& 
\frac{{\rm 2}}{{\rm 3}}\frac{{\rm \Theta } }{T}\left[g _{kl} 
+ \left({\rm 1} - {\rm 3} {\rm \Delta } \frac{\partial{p}}{\partial{\rho }} \right) 
u _{i}u _{k}\right] + \frac{{\rm 2} \sigma _{ik}}{T} \nonumber\\
&& - \frac{u _{k}}{T}\left[\dot{u}_{i} + \frac{\nabla  _{i}T}{T} \right] 
- \frac{u _{i}}{T}\left[\dot{u}_{k} + \frac{\nabla _{k}T}{T} \right] 
\ ,
\label{28}
\end{eqnarray}
where 
\begin{equation}
\Delta \equiv  1 - \frac{\Gamma }{\Theta } \ .
\label{29}
\end{equation}
We emphasize that the consistency requirement (\ref{25}) (with $\dot{s}=0$)  on which the expression (\ref{28})  is based only affects the projection 
$u ^{i}u ^{k}\pounds _{\xi } g _{ik}$ but not the parts involving  
$\dot{u} _{i}+\nabla  _{i}T/T$ and $\sigma _{ik}$. 
The introduction of the parameter $\Delta $ is convenient since it allows us 
to handle the cases of preserved and non-preserved particle numbers simultaneously. 
Particle number conservation $\Gamma =0$ is equivalent to $\Delta =1$. 
Particle production in Eq. (\ref{28}) and all further relations involving $\Delta $ is then straightforwardly taken into account by substituting 
$\Delta =1$ by the quantity (\ref{29}).   
As already mentioned, this formally simple replacement implies the transition 
from equilibrium configurations with vanishing entropy production to non-equilibrium situations with $S ^{a}_{;a}>0$.  
In the limit $\sigma _{ik}=0$, $\dot{u}_{i}+\nabla  _{i}T/T =0$, 
$\Delta =1$, and $\partial p/ \partial \rho =1/3$ Eq. (\ref{28})  reduces to the CKV equation for $u ^{i}/T$, i.e., to the familiar global equilibrium. 
In the following we briefly list up other 
previously investigated special cases of Eq. (\ref{28}):  

(i) 
A simple extension of the CKV condition is obtained by keeping  
$\sigma _{ik}=0$, $\dot{u}_{i}+\nabla  _{i}T/T =0$, 
$\Delta =1$, but allowing $\partial p/ \partial \rho$ to be different from $1/3$. 
The CKV property is modified in such a case, but there is no entropy production, i.e., the corresponding states are equilibrium states \cite{ZBGRG}. 
This represents an extension of the global equilibrium concept to 
equations of state different from $p= \rho /3$. 
Ultrarelativistic matter is then the limit in which the condition (\ref{28}) 
(with $\Delta =1$) reduces to the CKV condition for $\xi ^{a}$.     

(ii) A special case which implies a transition to non-equilibrium states corresponds to the vanishing of the parenthesis in front of $u _{i}u _{k}$ within the brackets in the first term on the right-hand side of Eq. (\ref{28}): 
\begin{equation}
3 \Delta \frac{\partial{p}}{\partial{\rho }} = 1 
\quad\Rightarrow\quad
\frac{\dot{T}}{T} = - \frac{\Theta }{3}\ . 
\label{30}
\end{equation}
If, moreover, $\dot{u}_{i}+\nabla  _{i}T/T =0$ and $\sigma _{ik}=0$  are valid, we have 
$\pounds _{\xi } g _{ik} = {\rm 2} \phi g _{ik} $, i.e.,  
$u ^{i}/T$ is a CKV. 
For $\Delta =1$, corresponding to $\Gamma =0$, i.e. particle number conservation,  the condition (\ref{30})  necessarily implies the ultrarelativistic equation of state $p=\rho /3$. 
For $\Gamma \neq 0$, however, equivalent to $\Delta \neq 1$, 
this condition is compatible with different equations of state 
as well \cite{ZTP,ZiBa1}.   
Admitting $\Gamma \neq 0$ corresponds to a 
non-equilibrium generalization of the 
global equilibrium.  
Again, ultrarelativistic matter is then no longer a singular case but just 
the limit for $\Gamma \rightarrow 0$.   

(iii) Another interesting case is $\Gamma =\Theta $ equivalent to 
$\Delta =0$, which corresponds to a stationary temperature behavior 
$\dot{T}=0$.    
For $\dot{u}_{i}+\nabla  _{i}T/T =0$ and $\sigma _{ik}=0$ the CKV condition 
$\pounds _{\xi } g _{ik} = {\rm 2} \phi g _{ik} $ is replaced by 
$\pounds _{\xi } g _{ik} = {\rm 2} \phi h _{ik} $ \cite{ZiBa2}. 
In Sec. IV the latter relation will be used to characterize a de Sitter phase 
in the early universe. 

It is obvious that there exist further generalizations of the global equilibrium which in general both modify the conformal symmetry and imply particle production, i.e. $\Delta \neq 1$.    
The CKV condition 
$\pounds _{\xi } g _{ik} = {\rm 2} \phi g _{ik} $ is known to follow from Boltzmann's equation as global equilibrium condition for the one-particle distribution function of a classical ultrarelativistic gas. 
For the special cases mentioned above it was shown that the corresponding modifications of the standard conformal symmetry with $p=\rho /3$ 
naturally appear as 
``generalized'' equilibrium conditions for the one-particle distribution function for particles which self-consistently move in additional force fields  \cite{ZiBa1,ZiBa2,ZBGRG}. 
In the following section  we extend these results to the general case (\ref{28}). 
 
\section{Kinetic theory in a force field}

\subsection{Generalized equlibrium conditions}

The particles of a relativistic gas are assumed to move under the influence of a 
four-force $F ^{i}$ in between elastic binary collisions, described by Boltzmann's collision integral $J \left[f,f \right]$. 
The equations of motion of the gas particles are  
\begin{equation}
\frac{\mbox{d} x ^{i}}{\mbox{d} \gamma  } = p ^{i}\ ,
\ \ \ \ \ \ 
\frac{\mbox{D} p ^{i}}{\mbox{d} \gamma  } = m F ^{i}\ ,
\label{31}
\end{equation}
where $\gamma $ is a parameter along the particle worldline which for massive particles 
may be related to the proper time $\tau $ by $\gamma   = \tau /m $. 
Since the particle four-momenta are normalized according to 
$p ^{i}p _{i} = -m ^{2}$, the force $F ^{i}$ has to satisfy the relation 
$p _{i}F ^{i} = 0 $. 
The corresponding equation for the invariant one-particle distribution function 
$ f = f\left(x\left(\tau  \right),p \left(\tau  \right)\right)$ may be written as 
\begin{equation}
\frac{\mbox{d}f}{\mbox{d} \tau} 
\equiv \frac{p ^{i}}{m }\left(\frac{\partial{}}{\partial{x ^{i}}} - \Gamma ^{k}_{il} p^{l}
\frac{\partial{}}{\partial{p^{k}}} \right) f
+ F^{i} \frac{\partial{f}}{\partial{p^{i}}}
=J [f,f]\ .
\label{32}
\end{equation}
Both the collision integral $J$ and the force $F ^{i}$ describe interactions within the many-particle system. 
While $J$ accounts for elastic binary collisions, 
$F ^{i}$ is to model different kinds of interactions in a simple manner. 
The force $F ^{i}$ is not a fundamental force. 
Strictly speaking, $F ^{i}$ should be calculated from the microscopic particle dynamics and, consequently, depend on the entire set of particle coordinates and momenta characterizing the system of gas particles.  
It is well known that except for simple cases a relativistic statistics for interacting many-particle systems is not available. 
In such a situation it seems natural to resort to a semi-phenomenological or effective theory. 
We assume that all the microscopic interactions, however involved their detailed structure may be, 
can be mapped onto an effective one-particle quantity, which may be used for a subsequent self-consistent treatment. 
The situation here is more complicated than for example for charged particles in an electromagnetic field were the microscopic starting point for a self-consistent analysis is a given interaction, the Lorentz force.  
Here, the analytic structure of the force is not known in advance. 
We restrict ourselves to the class of forces which admit  solutions 
of Eq. (\ref{32}) that are 
of the type of  J\"uttner's distribution function 
\begin{equation}
f^{0}\left(x, p\right) = 
\exp{\left[\alpha + \beta_{a}p^{a}\right] } \ ,
\label{33}
\end{equation}
where $\alpha = \alpha\left(x\right)$ and 
$\beta_{a}\left(x \right)$ is timelike. 
For $f \rightarrow f ^{0}$ the collision integral vanishes, i.e., 
$J\left[f ^{0},f ^{0}\right] = 0$. 
Substituting $f ^{0}$ into Eq. (\ref{32}) we obtain
\begin{equation}
p^{a}\alpha_{,a} +
\beta_{\left(a;b\right)}p^{a}p^{b}   
=  - m \beta _{i}F ^{i} 
\mbox{ .} 
\label{34}
\end{equation} 
Generally, the effective force $F ^{i}$ is an arbitrary function of the particle momentum. 
However, focusing on forces which satisfy the equilibrium condition (\ref{34})  implies considerable simplifications. 
This becomes obvious if we expand the quantity $\beta _{i}F ^{i}$ on the right-hand side of Eq. (\ref{34}) in a power series in $p ^{a}$: 
\begin{eqnarray}
\beta _{i}F ^{i} &=& \beta _{i}F _{0}^{i}\left(x \right) + 
\beta _{i}F ^{i}_{a}\left(x \right)p ^{a} 
+ \beta _{i}F ^{i}_{ab}\left(x \right)p ^{a}p ^{b} + ....\nonumber \\
&&+ \beta _{i}F ^{i}_{ab...n}\left(x \right)p ^{a}p ^{b}....p ^{n} + ...
\label{35}
\end{eqnarray}
Here we assume  $\beta _{i}F ^{i}_{a}$ also to contain all those trace parts of odd higher-order terms (third order and higher) which by the condition $p _{i}p ^{i} = - m ^{2}$ may be reduced to first order in $p ^{a}$. 
Analogously, $\beta _{i}F ^{i}_{ab}$ is supposed to contain all those trace parts of higher even order (fourth order and higher) terms which by the condition $p _{i}p ^{i} = - m ^{2}$ may be reduced to second order in $p ^{a}$.
This is equivalent to the assumption that the corresponding traces in 
$\beta _{i}F ^{i}_{ab...n}\left(x \right)p ^{a}p ^{b}....p ^{n}$ vanish for 
the third and all higher-order terms on the right-hand side of the expansion (\ref{35}). 
We emphasize that it is only the projection $\beta _{i}F ^{i}$ of 
$F ^{i}$ which is expanded in a power series, not the force  itself. For the latter there exists the additional requirement $p _{a}F ^{a}=0$. 

The point is now that the left-hand side of Eq. (\ref{34}) contains terms linear and quadratic in the particle momentum. 
Consequently, the right-hand side of this equation is restricted to quadratic order as well. 
It follows that the most general structure of the force term is 
\begin{equation}
\beta _{i}F ^{i} = 
\beta _{i}F ^{i}_{a}\left(x \right)p ^{a} 
+ \beta _{i}F ^{i}_{ab}\left(x \right)p ^{a}p ^{b} \ ,
\label{36}
\end{equation}
where we have used the redefinition 
\begin{equation}
F ^{i}_{kl} - \frac{F ^{i}_{0}}{m ^{2}}g _{kl} 
\rightarrow F^{i}_{kl} \ .
\label{37}
\end{equation}
Comparing now different powers in Eq. (\ref{34}) separately, we obtain the 
(generalized) equilibrium conditions
\begin{equation}
\alpha _{,a} = - m \beta _{i}F _{a}^{i}
\label{38}
\end{equation}
and 
\begin{equation}
\beta _{\left(k;l \right)} 
\equiv  \frac{1}{2}\pounds _{\beta }g _{kl} = - m \beta _{i}F^{i}_{kl}\ .
\label{39}
\end{equation}
The last equation relates the Lie derivative of the metric to the force exerted on the particle. 
With the identification $\beta _{i}\equiv  u _{i}/T$ we may read off the coefficient $u _{i}F ^{i}_{ab}$ which is necessary to recover condition (\ref{28}): 
\begin{eqnarray}
- m u _{i}F ^{i}_{ab} 
&=& \frac{\Theta }{3}\left[g _{ab} 
+ \left(1 - 3 \Delta \frac{\partial{p}}{\partial{\rho }}\right) 
u _{a}u _{b}\right] +  \sigma _{ab}\nonumber\\
&&
- \frac{u _{a}}{2}\left[\dot{u}_{b} 
+ \frac{\nabla _{b}T}{T} \right] 
- \frac{u _{b}}{2}\left[\dot{u}_{a} 
+ \frac{\nabla _{a}T}{T} \right]
\ .
\label{40}
\end{eqnarray}
This condition fixes the part which is quadratic in the particle momenta. 
The expression (\ref{40}) is the most general quadratic force coefficient 
in Eq. (\ref{36}) which is compatible with an equilibrium distribution function (\ref{33}).  
With $\alpha \equiv  \mu /T$ the linear part is determined by 
\begin{equation}
- \frac{m}{T} u _{i}F ^{i}_{a} 
= - \left(\frac{\mu }{T} \right)^{\displaystyle \cdot}u _{a} 
+ \nabla _{a} \left(\frac{\mu }{T} \right)\ .
\label{41}
\end{equation} 
An example for a force linear in $p ^{a}$ is the Lorentz force which is obtained for $F ^{ab}\rightarrow \frac{e}{m}F ^{ab}_{\left(em \right)}$  where $e$ is the charge and 
$F ^{ab}_{\left(em \right)}$ is the electromagnetic field strength tensor. 
For this case the equilibrium condition (\ref{38})  is well known in the literature 
\cite{Ehl,Groot}. Other cases have been discussed in a cosmological context \cite{ZiBa1,ZiBa2}. 

With Eqs. (\ref{39}) and (\ref{40}) we have recovered relation (\ref{28}) as equilibrium condition for a gas of particles moving in the field of a force (\ref{36}). 
It is a remarkable feature of this force that it depends on the kinematic quantities expansion, shear, acceleration and on the temperature gradient which are variables characterizing the system on a macroscopic level. 
These macroscopic quantities determine the motion of the individual microscopic particles which themselves are the constituents of the medium. 
In other words, our system is a self-interacting gas. 
The analogy to the well known self-consistent coupling of the particle momentum of a charged particle to the electromagnetic field strength tensor is obvious. 
The coefficient $u _{i}F ^{i}_{a}$  corresponding to the linear part of the force is determined by the spacetime behavior of the macroscopic variable 
$\mu /T$, which will be relevant for the energy-momentum balance to be discussed below.  

The relationship between Eq. (\ref{28}) and Eqs. (\ref{39}) and (\ref{40}) admits an alternative interpretation which exhibits features that are familiar from gauge field theories. 
Gauge field theories rely on the fact that local symmetry requirements (local gauge invariance) necessarily imply the existence of additional interaction fields (gauge fields). 
In the present context we impose 
the ``symmetry'' requirement (\ref{28}) (below we clarify in which sense the modification of the conformal symmetry is again a symmetry). 
Within the presented gas dynamical framework this ``symmetry'' can only be realized if one introduces additional interactions, here described by an effective force field $F ^{i}$. Consequently, in a sense, this force field may be regarded as the analogue of gauge fields. 

\subsection{Transport equations}

The particle number flow 4-vector 
$N^{i}$ and the energy momentum tensor $T^{ik}$ are
defined in a standard way (see, e.g., \cite{Ehl,IS}) as 
\begin{equation}
N^{i} = \int \mbox{d}Pp^{i}f\left(x,p\right) \mbox{ , } 
\ \ \ 
T^{ik} = \int \mbox{d}P p^{i}p^{k}f\left(x,p\right) \mbox{ .} 
\label{42}
\end{equation}
The integrals in the definitions (\ref{42}) and in the following  
are integrals over the entire mass shell 
$p^{i}p_{i} = - m^{2}$. 
The entropy flow vector $S^{a}$ is given by \cite{Ehl,IS} 
\begin{equation}
S^{a} = - \int p^{a}\left[
f\ln f - f\right]\mbox{d}P \mbox{ , }
\label{43}
\end{equation}
where we have restricted ourselves to the case of 
classical Maxwell-Boltzmann particles. 
The entropy production density is a sum of two terms: 
\begin{equation}
S ^{i}_{;i}  = \sigma _{c} + \sigma _{F}\ .
\label{44}
\end{equation}
Here, 
\begin{equation}
\sigma_{c} \equiv - \int \mbox{d}P 
J\left[f,f\right]  \ln f
\label{45}
\end{equation}
is the familiar contribution of Boltzmann's collision integral, while 
\begin{equation}
\sigma_{F} \equiv    m\int \mbox{d} P  
F ^{i}\frac{\partial{f}}{\partial{p ^{i}}}\ln f
\label{46}
\end{equation}
takes into account an entropy production due to the action of the force 
$F ^{i}$. 
Since Boltzmann's $H$ theorem guarantees $\sigma _{c} \geq 0$, 
we have 
\begin{equation}
S^i _{;i} - \sigma _{F} \geq 0 \ .
\label{47}
\end{equation}
The equality sign in the last relation, realized by $f \rightarrow f ^{0}$, characterizes the generalized equilibrium as a state with minimal entropy production.  
With $f$ replaced by $f^{0}$ in the definitions 
(\ref{42}) and (\ref{43}), the quantities $N^{a}$, $T^{ab}$ and $S^{a}$ may be 
split with respect to the unique 4-velocity $u^{a}$ according to 
\begin{equation}
N^{a} = nu^{a} \mbox{ , \ \ }
T^{ab} = \rho u^{a}u^{b} + p h^{ab} \mbox{ , \ \ }
S^{a} = nsu^{a} \mbox{  .}
\label{48}
\end{equation}
Here we have identified the general fluid quantities of the previous section with those following from the dynamics of a Maxwell-Boltzmann gas. 
Note that the energy-momentum tensor $T ^{ik}$ in Eqs. (\ref{42}) and (\ref{48})  does {\it not} coincide with the tensor 
$T ^{ik}_{\left(eff \right)}$ introduced in Eq. (\ref{1}).  
The exact integral expressions for $n$, $\rho$, $p$ are (see, e.g., \cite{Groot}), 
\begin{equation}
n =  \frac{p}{T} = \frac{4\pi m^{2}T}
{\left(2\pi\right)^{3}}K_{2}\left(
\frac{m}{T}\right) \exp{\left[\alpha \right]}\ ,
\label{49}
\end{equation}
and  
\begin{equation}
e =  \frac{\rho }{n} = m\frac{K_{1}\left(\frac{m}{T}\right)}
{K_{2}\left(\frac{m}{T}\right)} + 3 T = 
m\frac{K_{3}\left(\frac{m}{T}\right)}
{K_{2}\left(\frac{m}{T}\right)} -   T \ .
\label{50}
\end{equation}
The quantities $K_{n}$ are modified Bessel functions of the second
kind \cite{Groot}.  
The entropy per particle $s$ is given by 
\begin{equation}
s = \frac{\rho + p}{nT} - \frac{\mu }{T} \ .
\label{51}
\end{equation}
The balances  
\begin{eqnarray}
N^{a}_{;a}&=&-m \beta _{i}\int F ^{i}f ^{0}\mbox{d}P \ , \nonumber\\ 
T^{ak}_{\ ;k}&=&-m \beta _{i}\int p^{a}F ^{i}f ^{0}\mbox{d}P \ ,
\label{52}
\end{eqnarray}
may be written as 
\begin{equation}
N ^{a}_{;a} = - m \beta _{i}\left(F ^{i}_{a}N ^{a} 
+ F ^{i}_{ab}T ^{ab} \right)\ ,
\label{53}
\end{equation}
and 
\begin{equation}
T ^{ak}_{\ ;k} = - m \beta _{i}\left(F ^{i}_{b}T ^{ab} 
+ F ^{i}_{kl}M^{akl} \right)\ ,
\label{54}
\end{equation}
where $M ^{akl} = \int \mbox{d}P f ^{0}p ^{a}p ^{k}p ^{l}$ is the third moment of the equilibrium distribution function. 
For the entropy production density we find 
\begin{equation}
S^{a}_{;a} = m \beta _{i} \int \left[\alpha + \beta _{a}p ^{a} \right]
F ^{i}f ^{0}
\mbox{d}P 
=  - \alpha N^{a}_{;a} 
- \beta_{a}T^{ab}_{\ ;b}
\mbox{ . }
\label{55}
\end{equation}  
Using the expressions (\ref{40}) and (\ref{41})  and the decomposition (\ref{48}), 
the particle number balance (\ref{53}) becomes  
\begin{equation}
N ^{a}_{;a} = n \left(\frac{\mu }{T} \right)^{\displaystyle \cdot} 
+ n \Theta 
\left[\frac{p}{nT} - \frac{\rho }{nT}\Delta \frac{\partial{p}}{\partial{\rho }} \right]\ .
\label{56}
\end{equation}
From Eq. (\ref{16}) for $\dot{s}=0$ (which implies relation (\ref{26})) we obtain 
\begin{equation}
n \left(\frac{\mu }{T} \right)^{\displaystyle \cdot} = 
\frac{\rho + p}{T}\left(\frac{\partial{p}}{\partial{\rho }} 
- c _{s}^{2} \right)\Delta \Theta \ . 
\label{57}
\end{equation}
Since $p=nT$ the expression (\ref{15}) simplifies to   
\begin{eqnarray}
c _{s}^{2}  &=& \frac{nT}{\rho + p}
\left(1 + \frac{\partial{p}}{\partial{\rho }} \right)\nonumber\\
&&\quad\Rightarrow\quad
\left(\frac{\mu }{T} \right)^{\displaystyle \cdot} 
= \left[\frac{\rho }{nT}\frac{\partial{p}}{\partial{\rho }} 
- 1\right]\Delta \Theta \ .
\label{58}
\end{eqnarray}
Consequently, the balance (\ref{56}) reduces to  
\begin{equation}
N ^{a}_{;a} = 
\left(1 - \Delta  \right)n \Theta = n \Gamma \ ,
\label{59}
\end{equation}
which is consistent with Eq. (\ref{4}). 
For $\Delta = 1$ the particle number is conserved. 
To evaluate the energy-momentum balance (\ref{54}) 
one needs the relevant third moments of the distribution function, which may be obtained, e.g., with the help of the relation 
\begin{equation}
u _{a}p ^{a}f ^{0} = \frac{\partial{f ^{0}}}
{\partial \left({\frac{1}{T}} \right)}\ ,
\label{60}
\end{equation}
where the derivative has to be taken for $\alpha = {\rm const}$. 
One finds 
\begin{equation}
\frac{\partial{p}}{\partial{\frac{1}{T}}} = - T \left(\rho + p \right) \ ,
\mbox{\ \ \ }
\frac{\partial{\rho }}{\partial{\frac{1}{T}}} 
= - 3T \left(\rho + p \right) - pT \left(\frac{m}{T} \right)^{2}\ ,
\label{61}
\end{equation}
where we have used the properties 
\begin{equation}
\frac{\mbox{d}}{\mbox{d}z}\left(\frac{K _{n}\left(z \right)}{z ^{n}} \right) 
= - \frac{K _{n+1}\left(z \right)}{z ^{n}} \ ,
\mbox{\ \ \ }
K _{n+1} = 2n \frac{K _{n}}{z} + K _{n-1}
\label{62}
\end{equation}
of the functions $K _{n}$ \cite{Groot}. 
The relevant moments are 
\begin{equation}
g _{kl}M ^{akl} = g _{kl}\int \mbox{d}P f ^{0}p ^{a}p ^{k}p ^{l} 
= - n m ^{2} u ^{a}\ ,
\label{63}
\end{equation}
\begin{equation}
u _{a}u _{k}u_{l}M ^{akl} = - 3T \left(\rho + p \right) 
- p T \left(\frac{m}{T} \right)^{2}\ ,
\label{64}
\end{equation} 
and 
\begin{equation}
u _{k}\left(\dot{u}_{l} + \frac{\nabla  _{l}T}{T} \right)u _{a}M ^{akl} = 
\sigma _{kl}u _{a}M ^{akl} = 0 \ .
\label{65}
\end{equation}
For the energy balance we have 
\begin{eqnarray}
u _{a}T ^{ak}_{\  ;k} 
&=& - \rho \left(\frac{\mu }{T} \right)^{\displaystyle \cdot} 
- \left(\rho + p \right)\Theta
 \nonumber\\
&& + nT \Theta \Delta \frac{\partial{p}}{\partial{\rho }}
\left(3 \frac{\rho + p}{nT} + \left(\frac{m}{T} \right)^{2} \right)\ .
\label{66}
\end{eqnarray} 
With the help of the 
relation \cite{ZTP}
\begin{equation}
\left(\frac{m}{T} \right)^{2} - 1 + 5 \frac{\rho + p}{nT} 
- \left(\frac{\rho + p}{nT} \right)^{2} = 
\frac{\partial{\rho }}{\partial{p}}\ 
\label{67}
\end{equation} 
it follows that
\begin{equation}
u _{a}T^{ak}_{\  ;k} = - \left(1 - \Delta  \right)\Theta 
\left(\rho +  p\right)  \ .
\label{68}
\end{equation}
The right-hand side vanishes only for $\Delta = 1$. 
If we introduce the quantity $\Pi $ according to relations (\ref{26}) and  
(\ref{29}) we recover Eq. (\ref{5}).  
The relevant moments for the momentum balance 
$h _{ca}T^{ak}_{\  ;k}$ are 
\begin{equation}
u _{k}\left(\dot{u}_{l} + \frac{\nabla  _{l}T}{T} \right)
h _{ca}M ^{akl} = - \left(\dot{u}_{c} + \frac{\nabla  _{c}T}{T} \right) 
T \left(\rho + p \right)\ ,
\label{69}
\end{equation}
as well as $u _{k}u _{l}h _{ca}M ^{akl} = 0$, 
and  $ \sigma _{kl}h _{ca}M ^{akl} = 0$.  
From Eqs. (\ref{54}), (\ref{40}), (\ref{41}) and (\ref{48}) one obtains
\begin{eqnarray}
h _{ca}T^{ak}_{\  ;k} &\equiv&  \left(\rho + p \right)\dot{u}_{c} 
+ \nabla _{c}p \nonumber\\
&=& n T \nabla  _{c}\left(\frac{\mu }{T} \right) 
+ \left(\dot{u}_{c} + \frac{\nabla  _{c}T}{T} \right) 
\left(\rho + p \right)\ .
\label{70}
\end{eqnarray} 
In order to be consistent with Eq. (\ref{6}) we have to have 
\begin{equation}
nT \nabla  _{m}\left(\frac{\mu }{T} \right) 
+ \left(\dot{u}_{m} + \frac{\nabla  _{m}T}{T} \right)
\left(\rho + p \right) = 
- \nabla  _{m}\Pi  - \Pi  \dot{u}_{m}\ 
\label{71}
\end{equation}
with the {\it same} $\Pi  $ as above, given by Eq. (\ref{26}). 
Consequently, the quantity $\nabla _{a}\left(\mu /T \right)$ is determined by 
\begin{eqnarray}
n \nabla  _{m}\left(\frac{\mu }{T} \right) 
&=& - \frac{\rho + p}{T}\Delta 
\left(\dot{u}_{m} + \frac{\nabla  _{m}T}{T} \right) \nonumber\\
&&+ \nabla  _{m}\left[\left(1 - \Delta  \right)\frac{\rho + p}{T}
\right] \ .
\label{72}
\end{eqnarray}
This together with Eqs. (\ref{41}) and  (\ref{57}) determines the 
the most general linear part 
$-m u _{i}F ^{i}_{a}p ^{a}$ of the force, compatible with an equilibrium distribution function (\ref{33}).  

\subsection{Microscopic dynamics}

Once the effective one-particle force is known, we may study 
the motion of the gas particles explicitly. 
We extend here earlier investigations for special forces \cite{ZiBa1,ZiBa2,ZBGRG} and consider again the most general case. 
Contracting the equation of motion in Eq. (\ref{31}) with the macroscopic four-velocity results in 
\begin{equation}
\frac{\mbox{D}\left(u _{i}p ^{i} \right)}{\mbox{d}\tau } 
= u _{i}F ^{i} + \frac{1}{m}u _{i;k}p ^{i}p ^{k} \ ,
\label{73}
\end{equation}
where we have used that 
\[
\frac{\mbox{D} u ^{i}}{\mbox{d} \tau } = u ^{i}_{;n}\frac{p ^{n}}{m}\ .
\]
With 
the well-known decomposition of the covariant derivative of the 
four-velocity \cite{Ehlers,Ellis}, 
\begin{equation}
u _{i;n} = - \dot{u}_{i}u _{n} + \sigma _{in} +  \omega _{in}  
+ \frac{\Theta }{3}h _{in}\ ,
\label{74}
\end{equation}
where
$\omega_{ab} = h_{a}^{c}h_{b}^{d}u_{\left[c;d\right]}$, we find 
\begin{eqnarray}
\frac{\mbox{D}\left(u _{i}p ^{i} \right)}{\mbox{d}\tau } 
&=& u _{i}F ^{i} + \frac{1}{3m} \Theta h _{ik}p ^{i}p ^{k} \nonumber\\
&&+ \frac{1}{m}\sigma _{ik}p ^{i}p ^{k} 
- \frac{1}{m}\dot{u}_{i}u _{k}p ^{i}p ^{k}\ .
\label{75}
\end{eqnarray} 
The general structure of $u _{i}F ^{i}$ is given by 
Eqs. (\ref{36}), (\ref{40}) and 
(\ref{41}), 
\begin{eqnarray}
u _{i}F ^{i} & = & - \frac{T}{m}
\left[- \left(\frac{\mu }{T} \right)^{\displaystyle \cdot} u _{a} 
+ \nabla  _{a}\left(\frac{\mu }{T} \right)\right] p ^{a}  \nonumber\\
&&- \frac{1}{m}\left[\frac{\Theta }{3}
\left(g _{kl} + \left[1 - 3 \Delta \frac{\partial{p}}
{\partial{\rho }}\right] u _{k}u _{l}\right)  \right.
\nonumber \\ 
&&
- \left.\frac{1}{2}u _{k}\left(\dot{u}_{l}+ \frac{\nabla  _{l}T}{T } \right) \right.\nonumber\\
&&- \left.\frac{1}{2}u _{l}\left(\dot{u}_{k}+ \frac{\nabla _{k}T}{T } \right) 
+ \sigma _{kl}\right]p ^{k}p ^{l}\nonumber\\
& = & - \frac{T}{m}
\left[\left(\frac{\mu }{T} \right)^{\displaystyle \cdot} E 
+ p ^{a}\nabla  _{a}\left(\frac{\mu }{T} \right)\right]  \nonumber\\
&&- \frac{\Theta }{3m}
\left[\left(1 - 3 \Delta \frac{\partial{p}}{\partial{\rho }} \right) 
E ^{2} - m ^{2}\right] 
\nonumber \\ 
&&
- \frac{E}{m}p ^{i}
\left(\dot{u}_{i} + \frac{\nabla _{i}T}{T}\right) 
- \frac{\sigma _{ij}}{m}p ^{i}p ^{j}
\ ,
\label{76}
\end{eqnarray}
where $E \equiv  -u _{i}p ^{i}$ is the particle energy. 
Using the expression (\ref{76}) in Eq. (\ref{75})  together 
with Eq. (\ref{13}) for 
$\dot{s}=0$ yields 
\begin{eqnarray}
\frac{\mbox{d} E}{\mbox{d} \tau } 
&=& \frac{T}{m}E \left[\left(\frac{\mu }{T} \right)^{\displaystyle \cdot} 
+ \frac{E}{T}\frac{\dot{T}}{T}\right] \nonumber\\
&&+ \frac{T}{m}p ^{a}\left[\nabla _{a}\left(\frac{\mu }{T} \right) 
+ \frac{E}{T}\frac{\nabla _{a}T}{T}\right]\ .
\label{77}
\end{eqnarray}
Since 
\begin{equation}
\frac{\mbox{d}E}{\mbox{d}\tau } = \frac{E}{m}\dot{E} 
+ \left(p ^{a} - E u ^{a} \right)\frac{E _{,a}}{m}\ ,
\label{78}
\end{equation}
we obtain 
\begin{equation}
p ^{m}\left[\frac{\mu }{T} - \frac{E}{T} \right]_{,m} = 0 \ .
\label{79}
\end{equation}
This relation has to be valid for arbitrary $p ^{m}$. It follows that \begin{equation}
\alpha + \beta _{a}p ^{a}  =  \frac{\mu + u _{a}p ^{a}}{T} = 
- \frac{E - \mu }{T} =   {\rm const}\ ,
\label{80}
\end{equation}
i.e., the equilibrium distribution (\ref{33}) is indeed maintained without any restriction for the equation of state.  
This proves the consistency of our approach in the general case.

\section{The deflationary gas universe}

The cosmic medium is now regarded as a gas with internal self-interactions which macroscopically manifest themselves as an effective viscous pressure. 
For the discussion of the cosmological dynamics we 
restrict ourselves to a homogeneous and isotropic spatially flat universe.  Einstein's equations then reduce to   
\begin{equation}
8\pi G \rho = 3 H ^{2}\ ,
\mbox{\ \ \ \ }
\dot{H} = - 4\pi G\left(\rho + p + \Pi  \right)\ ,
\label{81}
\end{equation} 
where $H \equiv  \Theta /3 = \dot{a}/a$ is the Hubble rate and $a$ is the scale factor of the Robertson-Walker metric. 
Together with relation (\ref{26})  one finds 
\begin{equation}
\frac{\Gamma }{3H} = 1 + \frac{2}{3 \left(1+w \right) }
\frac{\dot{H}}{H ^{2}} 
\quad\Rightarrow\quad 
\Delta = - \frac{2}{3 \left(1+w \right) }\frac{\dot{H}}{H ^{2}}\ ,
\label{82}
\end{equation} 
where $w=p/ \rho $.  
According to Eqs. (\ref{26}), (\ref{29}), (\ref{36}), (\ref{40}), (\ref{41}), (\ref{58}), (\ref{72}), the viscous pressure is a consequence of the action of a force $F ^{i}$  on the microscopic constituents of the cosmic substratum with a projection 
\begin{eqnarray}
mu _{i}F ^{i} &=& \frac{2}{1+w}T
\left[\frac{1}{w}\frac{\partial{p}}{\partial{\rho }} - 1\right] 
\frac{\dot{H}}{H}E       \nonumber\\
&& - \left[E ^{2}\left(1+\frac{2}{1+w} 
\frac{\partial{p}}{\partial{\rho }}\frac{\dot{H}}{H ^{2}}\right) 
- m ^{2}\right]H \ .
\label{82a}
\end{eqnarray}
For the previously discussed special case in which $u ^{i}/T$  is a CKV with $w = \frac{T}{m}\ll 1$ and 
$\Delta = 1/2$, Eq. (\ref{82})  specifies to \cite{ZiBa1}
\begin{equation}
\frac{\dot{H}}{H ^{2}} = - \frac{3}{4} 
\quad\Rightarrow\quad H \propto \frac{4}{3t} 
\quad\Rightarrow\quad a \propto t ^{4/3}\ . 
\label{83}
\end{equation} 
We obtain accelerated expansion ($\ddot{a} > 0$) as a result of 
the back reaction of the production of massive particles on the Hubble parameter \cite{ZiBa1}. 
This dynamics is realized by a microscopic force with 
\begin{equation}
u _{i}F ^{i} \approx - \left(E-m \right)H \ .
\label{83a}
\end{equation}
A different special case is \cite{ZiBa2}
\begin{equation}
\pounds _{\xi }g _{ik} = \frac{{\rm 2}H}{T} h _{ik}\ .
\label{84}
\end{equation}
Applying here relation (\ref{21}) results in 
$\dot{T}/T = 0 \Rightarrow\ T = {\rm const}$. 
The force projection which gives rise to such a behavior is 
\begin{equation}
m u _{i}F ^{i} = - \left(E ^{2} - m ^{2} \right)H \ ,
\label{84a}
\end{equation}
equivalent to $\Delta =0$ or $\Gamma =3H$.   
According to Eq. (\ref{26}), a production rate $\Gamma = 3H $ is equivalent to an effective viscous pressure  $\Pi = - \left(\rho + p \right)$. 
Via Eq. (\ref{5})  this implies a constant energy density.  
Consequently, the case $\Gamma  = 3H$ corresponds to 
\begin{equation}
\frac{\dot{H}}{H ^{2}} = 0
\quad\Rightarrow\quad H = {\rm const} 
\quad\Rightarrow\quad  a \propto \exp{\left[Ht \right]}\ , 
\label{85}
\end{equation} 
i.e. exponential inflation.   
Both the CKV case on which the dynamics (\ref{83}) is based and the case (\ref{84}) correspond to constant ratios of 
$\Gamma / 3H $, i.e., to constant values of 
$\Delta $. 
Constant values of $\Delta $ (together with $w = {\rm const}$) give 
rise to a fixed expansion dynamics of the scale factor, (either a specific power law or exponential inflation for the specific choices above) which does not change in time. 
Since the early days of inflationary cosmology the graceful exit problem, i.e., 
the question of how to terminate a phase of accelerated expansion has played an essential role. 
A mechanism is necessary which induces a transition from accelerated to decelerated expansion. 
In the ``standard'' inflationary picture this mechanism is ``reheating'', a non-equilibrium period during which the inflaton field decays into ``ordinary'' 
matter through a complicated multi-level€ 
process with copious production of particles 
(see \cite{KLS1,KLS2,Boya,Brandenb}). 
Obviously, a setting in which $\Delta $ is assumed constant, does not provide us with a transition between accelerated and decelerated expansion. 
On this basis it is impossible to obtain a cosmological scenario with a changing effective equation of state for the cosmic medium. 
A constant $\Delta $ may be considered realistic at most piecewise.     
To study more general situations in wich $\Delta $ is allowed to depend on time 
we proceed as follows \cite{Z00}.  
We replace $\dot{H}$ in 
Eq. (\ref{82})  by  
\[
\dot{H} = \frac{\mbox{d}H}{\mbox{d}a}\dot{a} = H ^{\prime }H a \ ,
\]
where $H ^{\prime } \equiv  \mbox{d}H/ \mbox{d}a$. 
The resulting equation for the Hubble rate is 
\begin{equation}
\frac{H ^{\prime }}{\Delta H } 
= -\frac{3}{2}\frac{1+ w}{a}\ .
\label{86}
\end{equation}
For a constant $\Delta $ corresponding to 
$\Gamma \propto H$ it integrates to 
\begin{equation}
\frac{H}{H _{0}} 
= \left(\frac{a _{0}}{a}\right)^{\frac{3}{2}\left(1+w \right)\Delta }
\mbox{\ }\mbox{\ }\mbox{\ }
\mbox{\ }\mbox{\ }\mbox{\ }
\mbox{\ }\mbox{\ }\mbox{\ }
\mbox{\ }\mbox{\ }\mbox{\ }
\left(\Delta = {\rm const}\right)\ ,
\label{87}
\end{equation}
and 
\begin{equation}
a \propto t ^{\frac{2}{3 \left(1+w \right) \Delta }}
\mbox{\ }\mbox{\ }\mbox{\ }
\mbox{\ }\mbox{\ }\mbox{\ }
\mbox{\ }\mbox{\ }\mbox{\ }
\mbox{\ }\mbox{\ }\mbox{\ }
\left(\Delta = {\rm const} \right)\ .
\label{88}
\end{equation}
For $\Delta  = 1/2$ and $w=0$ we recover, of course, the behavior (\ref{83}). 
The case $\Delta =  0$ corresponds to exponential expansion.  
If the linear dependence of the particle production rate on the Hubble parameter is replaced by a quadratic one, i.e.,  $\Gamma \propto H ^{2}$, 
the quantity $\Delta $ will no longer be a constant. 
A dependence  of the type $\Gamma \propto H ^{2}$ was investigated in the context of matter creation due to  cosmological vacuum decay \cite{LiMa,GunzMaNe}. 
It is equivalent to a particle production which is proportional to the energy density. 
With the ansatz 
\begin{equation}
\frac{\Gamma }{3H} \propto  \frac{H}{H _{e}}\ ,
\label{89}
\end{equation}
where the quantity $H _{e} \equiv   H \left(a _{e} \right)$ is the Hubble parameter at some fixed epoch with $a = a _{e}$, integration of the differential equation (\ref{86}) for $w=1/3$  yields 
\begin{equation}
H = 2\frac{a _{e} ^{2}}
{a ^{2} + a _{e} ^{2}}H _{e}
\ .
\label{90}
\end{equation}
$H$ changes continously from $H = 2 H _{e}$, equivalent to 
$a \propto \exp{\left[Ht \right]}$ at $a=0$, to 
$H \propto a ^{-2}$, equivalent to $a \propto t ^{1/2}$, the standard radiation-dominated universe, for $a \gg a _{e}$. 
For $a<a _{e}$ we have $\ddot{a}>0$, while 
$\ddot{a}<0$ holds for $a>a _{e}$.  
The value $a _{e}$ denotes that transition from accelerated to decelerated expansion, corresponding to  
$\dot{H}_{e} = - H _{e}^{2}$. 
While a linear dependence of $\Gamma $ on $H$ corresponds to a power-law solution for the scale factor, we have now \cite{GunzMaNe}
\begin{equation}
t = t _{e} + \frac{1}{4H _{e}}
\left[\ln \left(\frac{a}{a _{e}} \right)^{2} + 
\left(\frac{a}{a _{e}} \right)^{2} - 1\right]\ .
\label{91}
\end{equation}
Our point here is to obtain a microscopic realization of such a scenario 
in terms of a self-interacting one-particle force of the type (\ref{82a}). 
From Eqs. (\ref{41}) and (\ref{58}) we find that for 
$p =nT$, $\rho =3nT$, which is the ultrarelativistic limit of the equation of state (\ref{50}), the linear part $u _{i}F ^{i}_{a}p ^{a}$ of the force projection vanishes. 
(Since the ultrarelativistic limit corresponds to $m \to 0$ we use here and in the following a refedinition of the force according to $m F^{i}\to F ^{i}$).  
With the Hubble rate (\ref{90}) we obtain that the quantity $\Delta $ is given by   
\begin{equation}
\Delta = \frac{a ^{2}}{a ^{2}+a _{e}^{2}}\ .
\label{92}
\end{equation}
For $a \rightarrow 0$ we have $\Delta \rightarrow 0$, i.e., 
$\Gamma \rightarrow 3H$,  while for $a \gg a _{e}$ we find 
$\Delta \rightarrow 1$, i.e., $\Gamma \rightarrow 0$. 
This allows us to obtain the corresponding quadratic force term. 
The result is 
\begin{equation}
-  u _{i}F ^{i} = \frac{a _{e}^{2}}{a ^{2}+a _{e}^{2}}E ^{2}H \ .
\label{92a}
\end{equation}
{\it An internal interaction, described by this component of an effective one-particle force $F ^{i}$ which is self-consistently exerted on the microscopic constituents of the cosmic medium, realizes the deflationary dynamics (\ref{90})}. 
Eqs. (\ref{39}) and (\ref{40}) specify to 
\begin{equation}
\pounds _{_{\frac{u _{a}}{T}}} g _{ik}  = 
\frac{{\rm 2} H}{T}
\left[g _{ik} + \frac{a _{e}^{{\rm 2}}}{a ^{{\rm 2}}
+a _{e}^{{\rm 2}}} u _{i}u _{k}\right]\ ,
\label{93}
\end{equation}
where $H$ is given by Eq. (\ref{90}).  
The limit $a \rightarrow 0$ corresponds to 
\begin{equation}
\pounds _{_{\frac{u _{a}}{T}}} g _{ik}  
= \frac{{\rm 2} H}{T}
h _{ik} 
\mbox{\ \ \ }\mbox{\ \ \ }\mbox{\ \ \ }\mbox{\ \ \ }\mbox{\ \ \ }
\mbox{\ \ \ }\mbox{\ \ \ }
\left(a \ll a _{e}\right)\ ,
\label{94}
\end{equation}
while for $a \gg a _{e}$ we reproduces the CKV condition 
\begin{equation}
\pounds _{_{\frac{u _{a}}{T}}} g _{ik} 
 = \frac{{\rm 2} H }{T}
g _{ik} 
\mbox{\ \ \ }\mbox{\ \ \ }\mbox{\ \ \ }\mbox{\ \ \ }\mbox{\ \ \ }
\mbox{\ \ \ }\mbox{\ \ \ }
\left(a \gg a _{e} \right)\ .
\label{95}
\end{equation} 
Formula (\ref{93}) continuosly mediates between an initial behavior (\ref{94})  and a final conformal symmetry (\ref{95}), equivalent to a change  between 
an initial phase with exponential expansion where $\Delta = 0$ and a final stage in which the universe is characterized by a standard global equilibrium  with $\Delta = 1$, i.e. $\Gamma = 0$,  for relativistic matter. 
Asymptotically, the vector $u ^{a}/T $  is a CKV of the metric $g _{ik}$.  
This brings us back to the question, to what extent a modification of the conformal symmetry (e.g. in Eq. (\ref{93})) represents a symmetry again. 
To solve this problem we introduce the ``optical metric'' (cf. \cite{Gordon,Ehlopt,Perlick})
\begin{equation}
\bar{g}_{ik} = g _{ik} + \left[1 - \frac{1}{n _{r}^{2}} \right]u _{i}u _{k}\ ,
\quad
n _{r}^{2} = 1 + \frac{a _{e}^{2}}{a ^{2}}= \Delta ^{-1}\ ,
\label{96}
\end{equation}
where $n _{r}$ plays the role of a refraction index of the medium which  
in the present case changes monotonically from $n _{r}\to \infty$ for 
$a \to 0$ to $n _{r} \to 1$ for $a \gg a _{e}$.  
For $a=a _{e}$ we have $n _{re}=\sqrt{2} $.  
Optical metrics are known to be helpful in simplifying the equations of light propagation in isotropic refractive media. 
With respect to optical metrics light propagates as in vacuum.  
Here we demonstrate that such type of metrics, in particular their symmetry properties, are also useful in relativistic gas dynamics. 
Namely, it is easy to realize that Eq. (\ref{93}) is equivalent to 
\begin{equation}
\pounds _{_{\frac{u _{a}}{T}}}\bar{g}_{ik} 
= \frac{{\rm 2}H}{T}\bar{g}_{ik} \ . 
\label{97}
\end{equation} 
{\it The quantity $u ^{a}/T$ is a CKV of the optical metric $\bar{g} _{ik}$.   
This clarifies in which sense the modification of the conformal symmetry is again a symmetry}. 
According to Eq. (\ref{96}) the introduction of an optical metric corresponds to a mapping of the non-equilibrium contributions ($\Delta \neq 1$) onto 
a refraction index $n _{r}\neq 1$ of the fluid. 
This mapping is such that it associates the same conformal symmetry to 
$\bar{g}_{ik}$ which characterizes $g _{ik}$ 
in the equilibrium case $\Delta =1$ with $n _{r}=1$ and 
$g _{ik}=\bar{g}_{ik}$. 
The  property of $u ^{a}/T$ being a CKV during the entire process allows us to precise the nature of the fluid dynamics which realizes the deflationary scenario (\ref{90}): 
The transition from a de Sitter phase to a FLRW period may be regarded as a specific non-equilibrium configuration which microscopically is characterized by an equilibrium distribution function and macroscopically by the conformal symmetry of an optical metric with a time dependent refraction index of the cosmic medium.

\section{Decaying vacuum model}

Combining the Friedmann equation (\ref{81}) with the Hubble parameter (\ref{90}), we obtain the energy density  
\begin{equation}
\rho = \frac{3 H _{e}^{2}}{2 \pi }m _{P}^{2}
\left[\frac{a _{e}^{2}}{a ^{2} + a _{e}^{2}} \right]^{2}\ ,
\label{98}
\end{equation}
where we have replaced $G$ by the Planck mass $m _{P}$  according to  
$G = 1 /m _{P}^{2}$. 
Initially, i.e., for $a \ll a _{e} $, corresponding to 
$t \rightarrow -\infty$, the energy density is constant, while for 
$a \gg a _{e}$ the familiar behaviour 
$\rho \propto a ^{-4}$ is recovered. 
The  temperature behaves as $T \propto \rho ^{1/4}$.  
The same dependence is obtained for the particle energy $E$, such that 
$E/T = {\rm const}$, consistent with Eq. (\ref{80}).   
The evolution of the universe starts in a quasistationary state with finite initial values of temperature and energy density at $a=0$ and approaches the standard 
radiation dominated universe for 
$a \gg a _{e}$. 
This scenario may alternatively be interpreted in the context of a decaying vacuum (cf. \cite{LiMa,GunzMaNe}). 
The energy density (\ref{98})  may be split into 
$\rho = \rho _{_{\left(v \right)}} + \rho _{_{\left(r \right)}}$ where  
\begin{eqnarray}
\rho _{_{\left(v \right)}} &=& \frac{3 H _{e}^{2}}{2 \pi }m _{P}^{2}
\left[\frac{a _{e}^{2}}{a ^{2} + a _{e}^{2}} \right]^{3}\ ,\nonumber\\
\rho _{_{\left(r \right)}} &=& \frac{3 H _{e}^{2}}{2 \pi }m _{P}^{2}
\left(\frac{a}{a _{e}} \right)^{2}
\left[\frac{a _{e}^{2}}{a ^{2} + a _{e}^{2}} \right]^{3}\ .
\label{99}
\end{eqnarray}
The part $\rho _{_{\left(v \right)}}$ is finite for $a \rightarrow 0$ and decays as $a ^{-6}$ for $a \gg a _{e}$, while 
the part $\rho _{_{\left(r \right)}}$ describes relativistic matter with 
$\rho _{_{\left(r \right)}}\rightarrow 0$ for $a \rightarrow 0$ and 
$\rho _{_{\left(r \right)}}\propto a ^{-4}$  for $a \gg a _{e}$. 
The energy balances are ($A = v, r$)
\begin{equation}
\dot{\rho }_{_{\left(A \right)}} + 3H \left[\rho _{_{\left(A \right)}} 
+ p _{_{\left(A \right)}}\right] 
= \Gamma _{_{\left(A \right)}}
\left[\rho _{_{\left(A \right)}} + p _{_{\left(A \right)}}\right]
\label{100}
\end{equation}
with
\begin{equation}
\frac{\Gamma _{_{\left(v \right)}}}{3H} 
= \left(1 - \frac{1}{2}\frac{a ^{2}}{a _{e}^{2}} \right)
\frac{a _{e}^{2}}{a ^{2} + a _{e}^{2}}\ ,
\mbox{\ \ \ }
\frac{\Gamma _{_{\left(r \right)}}}{3H} = \frac{3}{2}\frac{a _{e}^{2}}{a ^{2} + a _{e}^{2}}\ .
\label{101}
\end{equation}
The equation for $\rho _{_{\left(v \right)}}$ may be written as 
\begin{equation}
\dot{\rho} _{_{\left(v \right)}} 
+ 3H \left(\rho _{_{\left(v \right)}} 
+ p _{_{\left(v \right)}} + \Pi _{_{\left(v \right)}}\right) = 0 \ ,
\label{102}
\end{equation}
where
\begin{eqnarray}
\Pi _{_{\left(v \right)}} &\equiv&  
- \frac{\Gamma _{_{\left(v \right)}}}
{3H}\left(\rho _{_{\left(v \right)}} + p _{_{\left(v \right)}} \right) \nonumber\\
&=& - \left(1 - \frac{1}{2}\frac{a ^{2}}{a _{e}^{2}} \right)
\frac{a _{e}^{2}}{a ^{2} + a _{e}^{2}}
\left(\rho _{_{\left(v \right)}} + p _{_{\left(v \right)}} \right)\ .
\label{103}
\end{eqnarray}
This corresponds to an effective equation of state 
\begin{equation}
P _{_{\left(v \right)}} \equiv  p _{_{\left(v \right)}} 
+ \Pi _{_{\left(v \right)}} 
= \frac{a ^{2} - a _{e}^{2}}{a ^{2} + a _{e}^{2}}\rho _{_{\left(v \right)}}\ .
\label{104}
\end{equation}
Although we have always 
$p _{_{\left(v \right)}}= \rho _{_{\left(v \right)}}/3$, 
the effective equation of state for $a \rightarrow 0$ approaches 
$P _{_{\left(v \right)}} = - \rho _{_{\left(v \right)}}$. 
Effectively, this component behaves as a vacuum contribution. 
For $a \gg a _{e}$  it represents stiff matter with 
$P _{_{\left(v \right)}} = \rho _{_{\left(v \right)}}$.  
The radiation component may be regarded as emerging from the decay of the initial vacuum according to 
\begin{equation}
\dot{\rho }_{_{\left(r \right)}} + 4H \rho _{_{\left(r \right)}} 
= - \dot{\rho }_{_{\left(v \right)}}\ .
\label{105}
\end{equation}
One may introduce a radiation temperature $T _{_{\left(r \right)}}$ 
characterized by a dependence 
$T _{_{\left(r \right)}} \propto \rho _{_{\left(r \right)}}^{1/4}$.  
Similar thermodynamical consideration as those in sections II and III for the temperature $T$ are possible for $T _{_{\left(r \right)}}$. 
In particular, one may consider the Lie derivative of the metric $g _{ik}$  with respect to 
$\xi _{\left(r \right)}^{a}\equiv  u ^{a}/T _{_{\left(r \right)}}$. 
Restricting ourselves here to the isotropic case, we have, analogously to 
Eq. (\ref{20}), 
\begin{equation}
\pounds _{\xi _{\left(r \right)}}g _{ik} 
= {\rm 2} A _{_{\left(r \right)}}u _{i}u _{k} 
+ {\rm 2} \phi _{_{\left(r \right)}}h _{ik}\ ,
\label{106}
\end{equation}
where
\begin{equation}
A _{_{\left(r \right)}} = \frac{\dot{T}_{_{\left(r \right)}}}
{T ^{2}_{_{\left(r \right)}}}\ ,
\quad\quad 
\phi _{_{\left(r \right)}} 
= \frac{\dot{a}}{T _{_{\left(r \right)}}a}\ .
\label{107}
\end{equation}
Along similar lines as in the one-component case we obtain 
[cf. Eq. (\ref{93})]
\begin{eqnarray}
\pounds _{\xi _{\left(r \right)}} g _{ik}  
&\equiv&  \left(\frac{u _{i}}{T _{\left(r \right)}} \right)_{;k} 
+ \left( \frac{u _{k}}{T _{\left(r \right)}} \right)_{;i} \nonumber\\
&=& 
\frac{{\rm 2}H }{T _{\left(r \right)}}\left[g _{ik} 
+ \frac{{\rm 3}}{{\rm 2}} \frac{a _{{\rm e}}^{{\rm 2}}}{a ^{{\rm 2}} 
+ a _{{\rm e}}^{{\rm 2}}}
u _{i}u _{k}\right] 
\ .
\label{108}
\end{eqnarray}
The quantity $\xi ^{a}_{_{\left(r \right)}}$ is a CKV of the optical metric 
$\bar{g}_{ik}^{\left(r \right)}$,  
\begin{equation}
\pounds _{\xi _{\left(r \right)}}\bar{g}_{ik}^{\left(r \right)} 
= \frac{{\rm 2}H}{T _{_{\left(r \right)}}}\bar{g}_{ik}^{\left(r \right)} \ , 
\label{109}
\end{equation}
where
\begin{equation}
\bar{g}^{\left(r \right)}_{ik} = g _{ik} 
+ \left[1 - \frac{1}{ n_{r _{\left(r \right)}}^{2}} \right]u _{i}u _{k}\ ,
\label{110}
\end{equation}
with
\begin{equation}
n ^{2}_{r_{\left(r \right)}} = \left[1 + \frac{a _{e}^{2}}{a ^{2}} \right]^{3/2} 
= n _{r}^{3}\ .
\label{111}
\end{equation}
This demonstrates that the dynamics of the radiation component is governed by a separate conformal symmetry (\ref{109}) which, however, is closely related to the conformal symmetry (\ref{97}) for the medium as a whole. 
The corresponding optical metrics (\ref{110}) and (\ref{96}) have different effective refractive indices which are connected by Eq. (\ref{111}).  

\section{Conclusion}

We have presented a theory of self-interacting gaseous fluids which systematizes and generalizes earlier investigations concerning the relation between (non-)\linebreak equilibrium and symmetry properties of the cosmic substratum and the dynamics of the early universe. 
The general features of this approach are: 
(i) It provides an extension of the global equilibrium concept of relativistic gas dynamics. 
The standard global equilibrium for a simple gas which is possible only for ultrarelativistic matter is no longer a singular case but a well defined 
interaction-free limit of a more general concept. 
(ii) It admits a simultaneos treatment of equilibrium states and a specific class of non-equilibrium states. 
The latter are characterized by an equilibrium distribution function of 
the gas particles (generalized equilibrium) although they may be far from equilibrium.  

We have established a microscopic realization of a deflationary scenario of the early universe according to which the cosmological evolution  started with a  de Sitter period, followed by a smooth transition to a standard radiation dominated FLRW phase. 
This deflationary dynamics represents a specific, exactly solvable  non-equilibrium configuration of a self-interacting gas. 
Mapping the non-equilibrium contributions onto an effective refraction index of the cosmic matter, the deflationary transition appears as  
the manifestation of a timelike conformal symmetry of an optical metric in which the refraction index changes smoothly from a very large value in the de Sitter period to unity in the FRLW phase.

\acknowledgments
This paper was supported by the Deutsche Forschungsgemeinschaft.

\end{document}